\documentstyle[11pt,newpasp,twoside,epsf,natbib]{article}

\def\edcomment#1{\iffalse\marginpar{\raggedright\sl#1\/}\else\relax\fi}
\reversemarginpar

\begin{document}
\title{Mergers and Non-Thermal Processes in Clusters}
 \author{Craig L. Sarazin}
\affil{Department of Astronomy, University of Virginia,
P. O. Box 3818, Charlottesville, VA 22903-0818 USA}

\begin{abstract}
Clusters of galaxies generally form by the gravitational merger of
smaller clusters and groups.
Mergers drive shocks in the
intracluster gas which heat the intracluster gas.
Mergers disrupt cluster cooling cores.
Mergers produce large, temporary increases in the X-ray luminosities and
temperatures of cluster; such merger boost may bias estimates of
cosmological parameters from clusters.
Chandra observations of the X-ray signatures of mergers, particularly
``cold fronts,'' will be discussed.
X-ray observations of shocks can be used to determine the 
kinematics of the merger.
As a result of particle acceleration in
shocks and turbulent acceleration following mergers, clusters of galaxies
should contain very large populations of relativistic electrons and ions.
Observations and models for the radio, extreme ultraviolet, hard X-ray,
and gamma-ray emission from nonthermal particles accelerated in these
shocks are described.
\end{abstract}

\section{Introduction}
\label{sec:sarazin_intro}

Major cluster mergers are the most energetic events in the Universe
since the Big Bang.
In these mergers, the subclusters collide at velocities of
$\sim$2000 km/s,
releasing gravitational binding energies of as much as $\ga$$10^{64}$
ergs.
Figure~1a shows the Chandra image of the merging cluster
Abell~85, which has two subclusters merging with the main
cluster
(Kempner, Sarazin, \& Ricker 2002).
The relative motions in mergers are moderately supersonic,
and shocks are driven into the intracluster medium.
In major mergers, these hydrodynamical shocks dissipate energies of
$\sim 3 \times 10^{63}$ ergs; such shocks are the major heating
source for the X-ray emitting intracluster medium.
Mergers shocks heat and compress the X-ray emitting intracluster
gas, and increase its entropy.
We also expect that particle acceleration by these shocks will produce
nonthermal electrons and ions, and these can produce synchrotron
radio, inverse Compton (IC) EUV and hard X-ray, and gamma-ray emission.

\begin{figure}
\plotfiddle{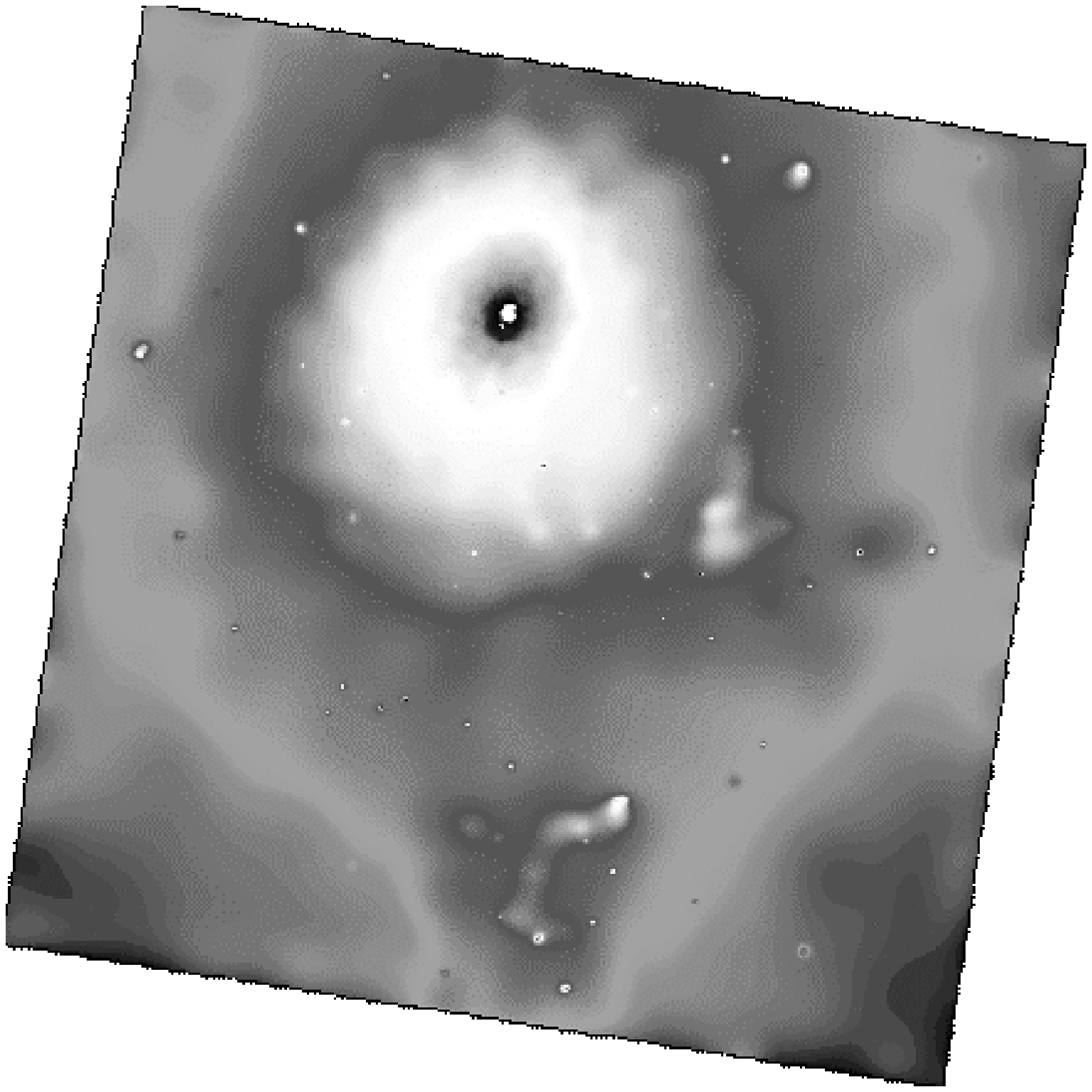}{1.8truein}{0}{35}{35}{-165}{-10}
\vskip-2.0truein
\plotfiddle{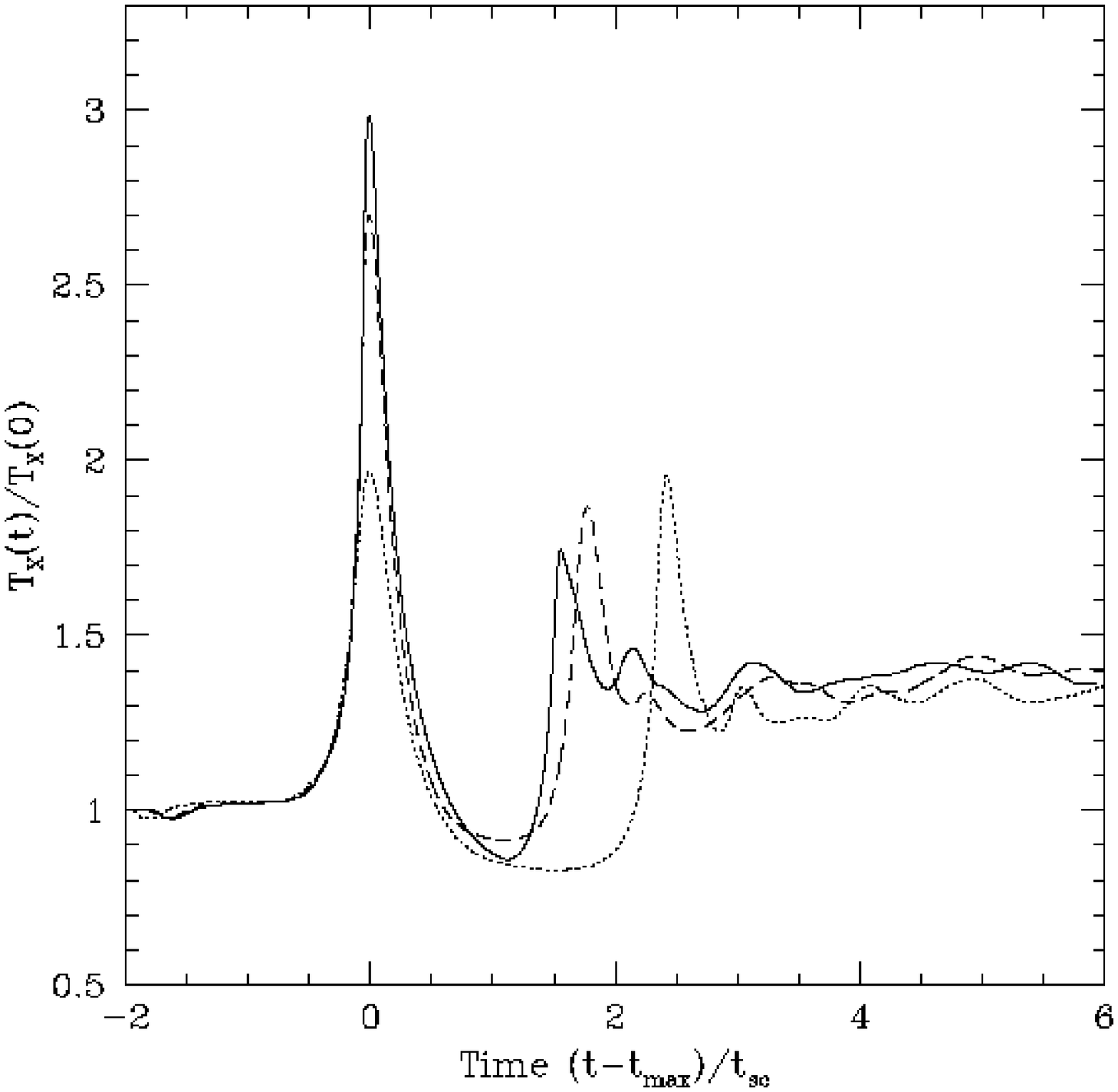}{1.8truein}{0}{28}{28}{8}{-50}
\caption{
\label{fig:A85+1E}
(a) The Chandra X-ray image of the merging cluster Abell~85
(Kempner et al.\ 2002).
Two subclusters to the south and southwest are merging with the main cluster.
(b)
The X-ray emission-averaged temperature in a
pair of equal mass clusters undergoing a merger
(Ricker \& Sarazin 2001;
Randall et al.\ 2002)
}
\end{figure}

\section{Thermal Effects of Mergers}
\label{sec:sarazin_thermal}

Mergers heat and compress the intracluster medium.
Shocks associated with mergers also increase the entropy of the
gas.
Mergers can help to mix the intracluster gas, possibly removing abundance
gradients.
Mergers appear to disrupt the cooling cores found in many clusters;
there is an anticorrelation between cooling core clusters and clusters
with evidence for strong ongoing mergers
(e.g., Buote \& Tsai 1996).
The specific mechanism by which cooling cores are disrupted is not
completely understood at this time
(e.g., Ricker \& Sarazin 2001).

The heating and compression associated with mergers can produce a
large, temporary increase in
the X-ray luminosity (up to a factor of $\sim$10) and
the X-ray temperature (up to a factor of $\sim$3) of the merging clusters
(Figure 1b; Ricker \& Sarazin 2001;
Randall, Sarazin, \& Ricker 2002).
Very luminous hot clusters are very rare objects in the Universe.
Although major mergers are also rare events, merger boosts can cause
mergers to strongly affect the statistics of the most luminous, hottest
clusters.
Simulations predict that many of the most luminous, hottest clusters
are actually merging systems, with lower total masses than would be
inferred from their X-ray luminosities and temperatures
(Randall et al.\ 2002).
Since the most massive clusters give the greatest leverage in determining
$\Omega_M$ and $\sigma_8$, these values can be biased by merger boosts.

One of the most dramatic results on clusters of galaxies to come from
the Chandra X-ray observatory was the discovery of sharp surface brightness
discontinuities in the images of merging clusters
(Figure 1a).
These were first seen in Abell 2142
(Markevitch et al.\ 2000)
and Abell 3667
(Vikhlinin, Markevitch, \& Murray 2001b).
Initially, one might have suspected these features were merger shocks,
but X-ray spectral measurements showed that the dense, X-ray bright
``post-shock'' gas was cooler, had lower entropy, and was at the same
pressure as the lower density ``pre-shock'' gas.
This would be impossible for a shock.
Instead, these ``cold fronts'' are apparently contact discontinuities between
gas which was in the cool core of one of the merging subclusters and
surrounding shocked intracluster gas
(Vikhlinin et al.\ 2001b).
The cool cores are plowing rapidly through the shocked cluster gas,
and ram pressure sweeps back the gas at the front edge of the cold front.
In a few cases (e.g., 1E0657-56; Markevitch et al.\ 2003), bow shocks are
seen ahead of the cold fronts.

\begin{figure}
\vskip-0.1truein
\plotfiddle{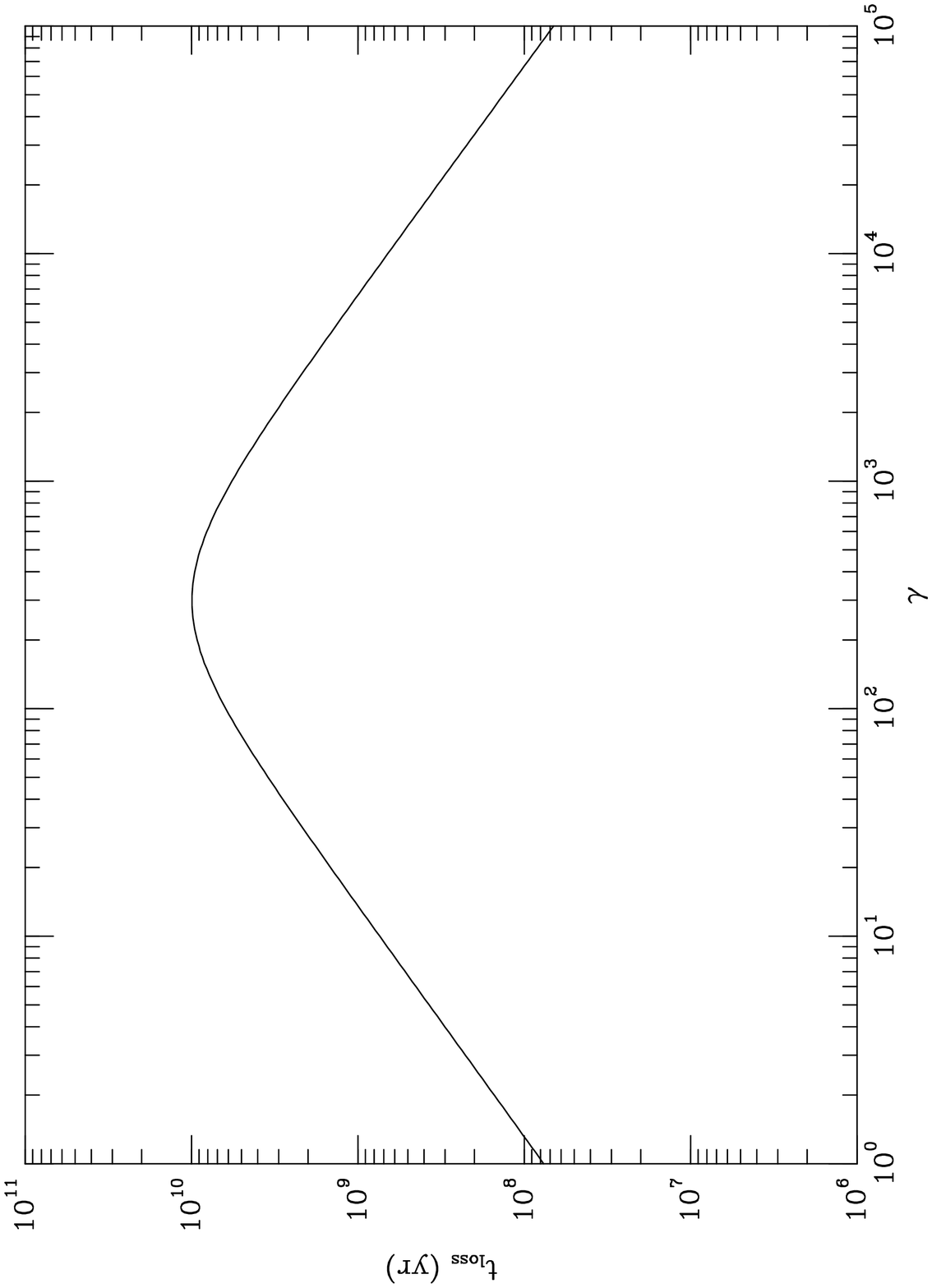}{1.6truein}{-90}{25}{25}{-200}{125}
\vskip-1.8truein
\plotfiddle{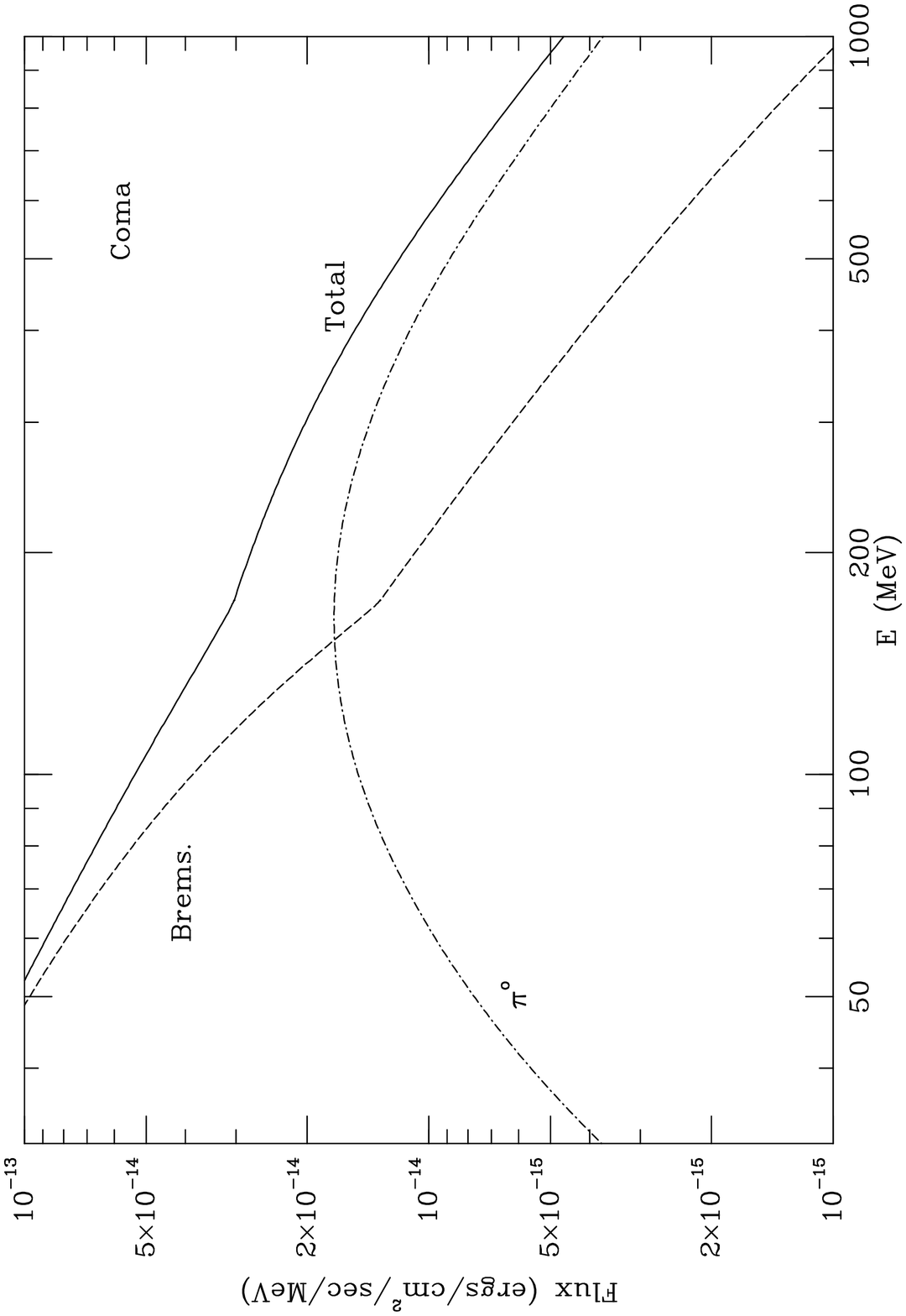}{1.6truein}{-90}{25}{25}{-10}{125}
\caption{
\label{fig:lifetime+gamma}
(a) The lifetimes of relativistic electrons in a typical cluster as
a function of their Lorentz factor $\gamma$
(Sarazin 1999a).
(b)
The gamma-ray emission spectrum from a model for relativistic particles in
the Coma cluster
(Sarazin 1999b).
Emission from electrons (Brems.) and ions (due to $\pi^o$ decays) are
shown separately.
}
\end{figure}

Cold fronts and merger shocks provide a number of classical hydrodynamical
diagnostics which can be used to determine the kinematics of the merger
(Vikhlinin et al.\ 2001b;
Sarazin 2002).
Most of these diagnostics give the merger Mach number $\cal M$.
The standard Rankine-Hugoniot shock jump conditions can be applied to
merger shocks; for example, the pressure discontinuity is
\begin{equation}
\frac{P_2}{P_1} = \frac{ 2 \gamma}{\gamma + 1} {\cal M}^2 -
\frac{\gamma - 1}{\gamma + 1} \, ,
\label{eq:sarazin_shock}
\end{equation}
where $P_1$ and $P_2$ are the pre- and post-shock pressure, and
$\gamma = 5/3$ is the adiabatic index of the gas.
For bow shocks in front of cold front, the shock may be conical at the
Mach angle, $\theta_m = \csc^{-1} ( {\cal M}_1 )$.
The ratio of the pressure at the stagnation point in front of a cold front
to the pressure well ahead of it is given by
\begin{equation}
\frac{P_{\rm st}}{P_1} = \left\{
\begin{array}{cl}
\left( 1 + \frac{\gamma_ - 1}{2} {\cal M}^2
\right)^{\frac{\gamma}{\gamma - 1}} \, , &
{\cal M} \le 1 \, , \\
{\cal M}^2 \,
\left( \frac{\gamma + 1}{2}
\right)^{\frac{\gamma + 1}{\gamma - 1}} \,
\left( \gamma - \frac{\gamma - 1}{2 {\cal M}^2} \,
\right)^{- \frac{1}{\gamma - 1}} \, , &
{\cal M} > 1 \, . \\
\end{array}
\right.
\label{eq:sarazin_stagnation}
\end{equation}
If the motion of the cold front is supersonic, the stand-off distance
between the cold front and the bow shock varies inversely with $\cal M$.
For major mergers, these kinematic diagnostics generally indicate that
mergers are mildly transonic ${\cal M} \approx 2$, corresponding to
merger velocities of $\sim$2000 km/s.

Mergers also provide a useful environment for testing the role of various
physical processes in clusters.
For example, the very steep temperature gradients at cold fronts imply that
thermal conduction is suppressed by a large factor ($\sim$$10^2$;
Ettori \& Fabian 2000;
Vikhlinin, Markevitch, \& Murray 2001a),
presumably by magnetic fields.
The smooth front surfaces of some cold fronts suggest that Kelvin-Helmholtz
instabilities are being suppressed, also probably by magnetic fields.
Recently, Markevitch et al.\ (2003) have used the relative distributions of
dark matter, galaxies, and gas in the dramatic merging cluster
1E0657-56 (Figure 1b) to argue that the collision cross-section per unit
mass of the dark matter must be low, $\sigma/m \la 1$ cm$^2$/g, which
excludes most of the self-interacting dark matter models invoked to explain
the mass profiles of galaxies.

\section{Non-Thermal Effects of Mergers}
\label{sec:sarazin_nonthermal}

High speed astrophysical shocks in diffuse gas generally lead to
significant acceleration of relativistic electrons.
For example, typical supernova remnants have blast wave shock velocities of
a few thousand km/s, which are comparable to the speeds in merger shocks.
(However, the Mach numbers in merger shocks are much lower.)
The ubiquity of radio emission from Galactic supernova remnants implies
that at least a few percent of the shock energy goes into accelerating
relativistic electrons, with more probably going into ions.
If these numbers are applied to strong merger shocks in clusters, one would
expect that relativistic electrons with a total energy of
$E_{\rm rel,e} \sim 10^{62}$ erg would be accelerated, with even more energy
in the relativistic ions.
Thus, merging clusters should have huge populations of relativistic
particles.

\begin{figure}
\vskip-0.15truein
\plotfiddle{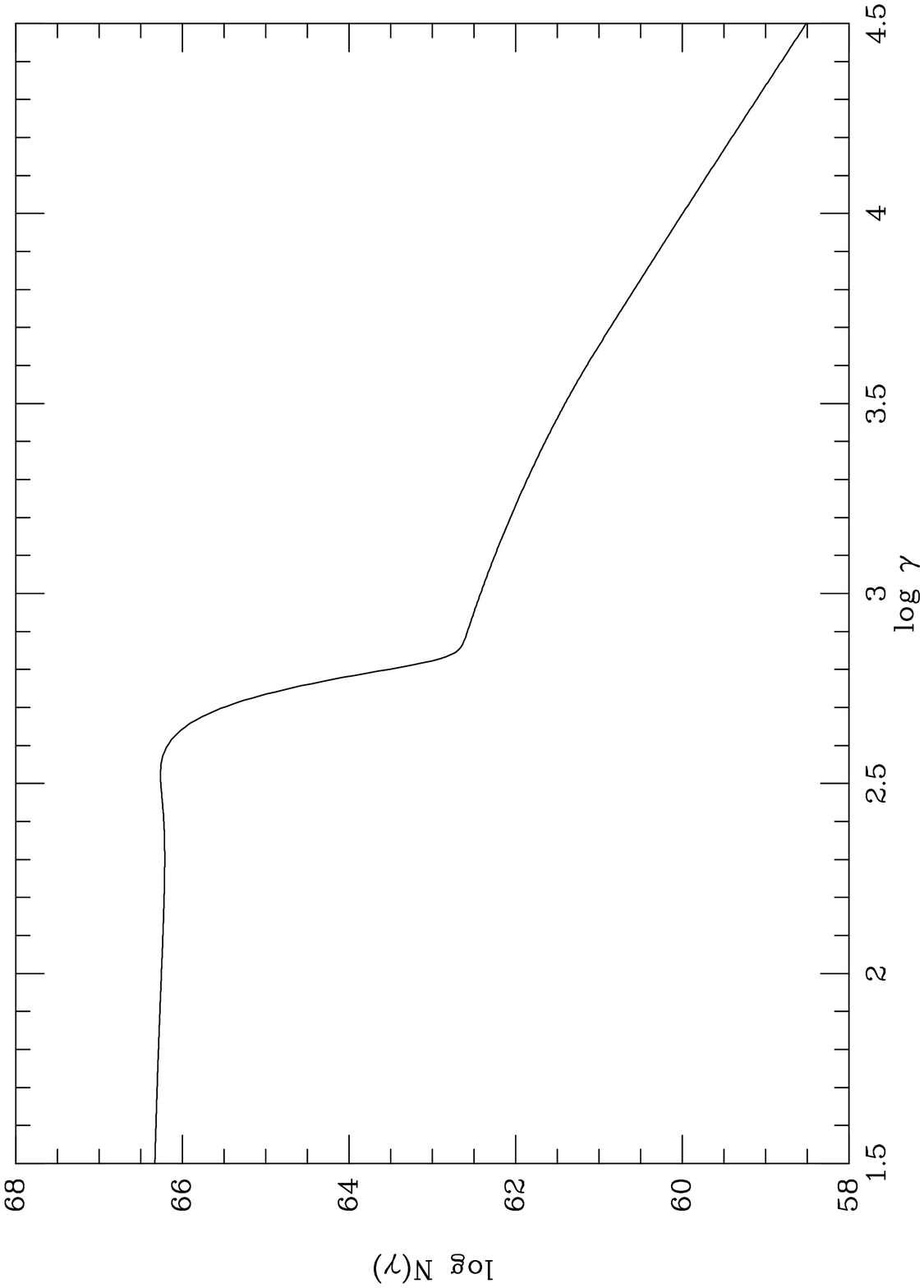}{1.6truein}{-90}{25}{25}{-200}{125}
\vskip-1.8truein
\plotfiddle{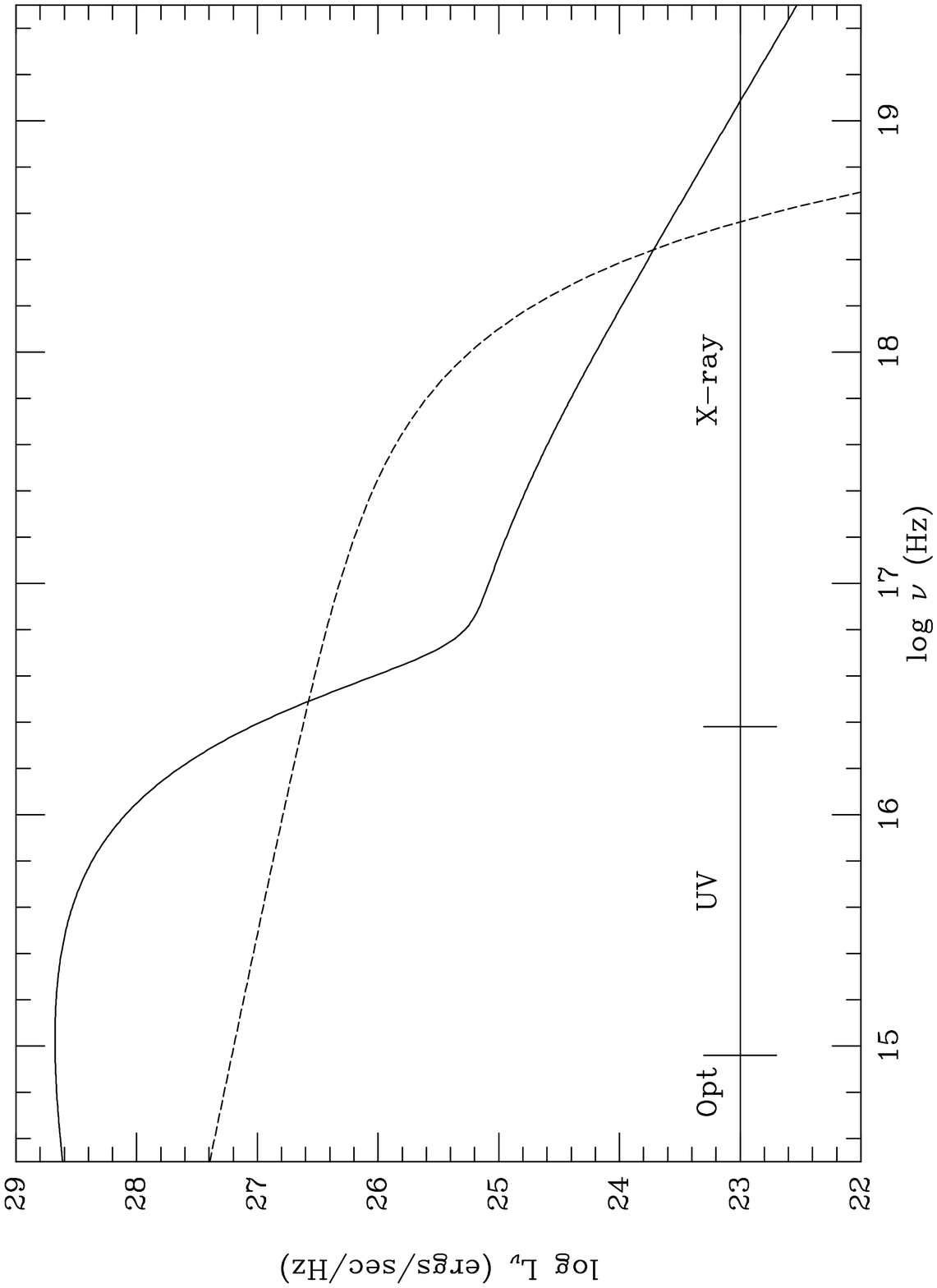}{1.6truein}{-90}{25}{25}{-10}{125}
\caption{
\label{fig:espect+ic}
(a) The energy spectrum of relativistic electrons in a model for
a merging cluster
(Sarazin 1999a).
The large population at $\gamma \sim 300$ are due to many previous mergers,
while the tail to high energies is due to the current merger.
(b) The IC spectrum produced by the model in (a).
For reference, the dashed line is thermal bremsstrahlung at a typical
cluster luminosity.
}
\end{figure}

Clusters should also retain some of these particles for very long times.
The cosmic rays gyrate around magnetic field lines, which are frozen-in to
the gas, which is held in by the strong gravitational fields of clusters.
Because clusters are large, the timescales for diffusion are generally
longer than the Hubble time.
The low gas and radiation densities in the intracluster medium imply that
losses by relativistic ions are very slow, and those by relativistic
electrons are fairly slow.
Figure 2a shows the loss timescale for electrons under typical cluster
conditions;
electrons with Lorentz factors $\gamma \approx 300$ and energies of
$\approx$150 MeV have lifetimes which approach the Hubble time, as long
as cluster magnetic fields are not too large ($B \la 3$ $\mu$G).
As a result, clusters should contain two populations of primary relativistic
electrons (Figure 3a):
those at $\gamma \sim 300$ which have been produced by mergers
over the lifetime of the clusters; and a tail to higher energies produced
by any current merger.

The lower energy electrons will mainly be visible in the EUV/soft X-ray
range (Figure 3b).
Such emission has been seen in clusters, although its origin is uncertain
and may well be thermal, as Jonathon Mittaz described in this session.
More energetic electrons, with energies of many GeV, produce hard X-ray
IC emission and radio synchrotron emission.
Diffuse radio sources, not associated with radio galaxies, have been
observed for many years in merging clusters.
Centrally located, unpolarized, regular sources are called ``radio halos'',
while peripheral, irregular, polarized sources are called ``radio
relics''.
Recent Chandra observations seem to show a direct connection between
radio halos and merger shocks in clusters
(Markevitch \& Vikhlinin 2001).
The same higher energy electrons will produce hard X-ray IC emission.
Recently, such emission appears to have been detected with BeppoSAX
and RXTE, although the detections are relatively weak and controversial
[see Rossetti \& Molendi (2003) and the poster by Fusco-Femiano et al.\
in this session].

I believe one of the most exciting possibilities for the future is the
detection of clusters in hard gamma-ray radiation
(Figure 2b).
Essentially, all models for the nonthermal populations in clusters predict
that they should be very luminous gamma-ray sources, particularly at photon
energies of $\sim$100 MeV
(Sarazin 1999b;
Blasi \& Gabici 2003).
The emission at these energies is partly due to electrons with energies
of ~150 MeV, which should be ubiquitous in clusters.
One nice feature of this spectral region is that emission is produced both
by relativistic electrons (bremsstrahlung and IC) and relativistic ions
(through the production of $\pi^o$ particles which decay into gamma-ray
photons; see Figure 2b).
Thus, one can determine both population is clusters.
Models suggest that GLAST and AGILE will detect $\sim$40 nearby clusters.

\acknowledgments
This work was supported by the National
Aeronautics and Space Administration through
Chandra Award Numbers GO1-2123X, GO2-3159X, and GO3-4160X, and
{\it XMM/Newton} Award Numbers
NAG5-10075 and NAG5-13088.

\end{document}